\pdfoutput=1
\documentclass{article}
\usepackage{frascatiphys,graphicx,subfigure}
\providecommand{\bar}[1]{\overline{#1}}
\newcommand{\kaon}{\mathrm{K}^0}
\newcommand{\akaon}{\bar{\mathrm{K}^0}}

\newcommand{\Ks}{\mathrm{K_S}}
\newcommand{\Kl}{\mathrm{K_L}}

\newcommand{\Kp}{\mathrm{K_{+}}}
\newcommand{\Km}{\mathrm{K_{-}}}
\usepackage{amsmath, amssymb}
\usepackage{braket}
\usepackage{tikz}
\begin{document}
\title{\textbf{A direct test of $\mathcal{T}$ symmetry in the neutral K meson system with $\Ks\to\pi \ell \nu$ and $\Kl\to3\pi^0$ at KLOE-2}
}
\author{
Aleksander Gajos\\on behalf of the KLOE-2 Collaboration\\
{\em Institute of Physics, Jagiellonian University, Cracow, Poland} 
\\
}
\maketitle
\baselineskip=11.6pt
\begin{abstract}
  Quantum entanglement of K and B mesons allows for a direct experimental test of time-reversal symmetry independent of $\mathcal{CP}$ violation. The $\mathcal{T}$ symmetry can be probed by exchange of initial and final states in the reversible transitions between flavor and CP-definite states of the mesons which are only connected by the $\mathcal{T}$ conjugation. While such a test was successfully performed by the BaBar experiment with neutral B mesons, the KLOE-2 detector can probe $\mathcal{T}$-violation in the neutral kaons system by investigating the process with $K_S\to \pi^{\pm} l^{\mp} \nu_{l}$ and $K_L\to3\pi^0$ decays. Analysis of the latter is facilitated by a novel reconstruction method for the vertex of $K_L\to3\pi^0$ decay which only involves neutral particles. Details of this new vertex reconstruction technique are presented as well as prospects for conducting the direct $\mathcal{T}$ symmetry test at the KLOE-2 experiment.
\end{abstract}
\baselineskip=14pt
\section{Introduction}
A direct test of the time-reversal symmetry in a single experiment is of great interest among possible ways to probe the $\mathcal{T}$ symmetry violation\cite{Wolfenstein:1999re}. For particles with spin 0 such as pseudo-scalar mesons, a direct test may be obtained by observation of an asymmetry between a reaction from state \textit{i} to state \textit{f} and a reversed reaction $f\to i$. While the CPLEAR experiment measured a nonzero value of the Kabir asymmetry in neutral kaon oscillations\cite{Angelopoulos:1998dv}, a controversy was raised as to whether this result was independent of $\mathcal{CP}$ violation as the $\kaon \to \akaon$ and $\akaon \to \kaon$ transitions are connected by both the $\mathcal{T}$ and $\mathcal{CP}$ symmetries. Therefore, an idea was proposed to exploit the quantum correlations of neutral B and K meson pairs to observe reversible transitions between flavour and $\mathcal{CP}$-definite states of the mesons\cite{Bernabeu:2012ab, Bernabeu:2012nu}. Such a $\mathcal{T}$ symmetry test was successfully performed by the BaBar experiment with the entangled neutral B meson system\cite{Lees:2012uka}. In turn, the KLOE-2 detector at the DA$\Phi$NE $\phi$-factory is capable of performing a statistically significant direct observation of $\mathcal{T}$ symmetry violation with neutral kaons independently of $\mathcal{CP}$ violation\cite{Bernabeu:2012nu}.
\section{Transitions between flavour and $\mathcal{CP}$-definite neutral kaon states}
Neutral kaon states may be described in a number of bases including flavour-definite states:
\begin{equation}
  \mathcal{S} \ket{\kaon}  = +1\ket{\kaon}, \qquad
  \mathcal{S} \ket{\akaon} = -1\ket{\akaon},
\end{equation}
as well as the states with definite $\mathcal{CP}$ parity:
\begin{eqnarray}
  \ket{\Kp} & =\frac{1}{\sqrt{2}}\left[\ket{\kaon}+\ket{\akaon}\right]\qquad \mathcal{CP}=+1,\\
  \ket{\Km} & =\frac{1}{\sqrt{2}}\left[\ket{\kaon}-\ket{\akaon}\right]\qquad \mathcal{CP}=-1.
\end{eqnarray}
State of the kaon can be identified at the moment of decay through observation of the decay final state. With the assumption of $\Delta S=\Delta Q$ rule\footnote{Althought an assumption, the $\Delta S=\Delta Q$ rule is well tested in semileptonic kaon decays\cite{Beringer:1900zz}}, semileptonic kaon decays with positively and negatively charged leptons (later denoted as $\ell^+$, $\ell^-$) unambiguously identify the decaying state as $\kaon$ and $\akaon$ respectively. Similarly, the $\mathcal{CP}$-definite states $\Kp$ and $\Km$ are implied by decays to hadronic final states with respectively two and three pions (denoted $\pi\pi$, $3\pi$).
In order to observe a transition between the $\{\kaon, \akaon\}$ and $\{\Kp, \Km\}$ states, both the \textit{in} and \textit{out} states must be identified in the respective basis. This is uniquely possible in the entangled system of neutral K mesons produced at a $\phi$-factory. Due to conservation of $\phi(1^{--})$ quantum numbers, the $\phi\to\kaon\akaon$ decay yields an anti-symmetric non-strange final state of the form:
\begin{equation}
    \Ket{\phi} \to \frac{1}{\sqrt{2}} \left( \Ket{\kaon(+\vec{p})}\Ket{\akaon(-\vec{p})}-\Ket{\akaon(+\vec{p})}\Ket{\kaon(-\vec{p})} \right),
\end{equation}
which exhibits quantum entanglement between the two kaons in the EPR sense\cite{Einstein:1935rr}. Thus, at the moment of decay of first of the K mesons (and, consequently, identification of its state) state of the partner kaon is immediately known to be orthogonal. This property allows for identification of state of the still-living kaon only by observing the decay of its partner. Its state can be then measured at the moment of decay after time~$\Delta t$, possibly leading to observation of a transition between strangeness and CP-definite states. A list of all possible transitions is presented in Table \ref{ag:tab:states}. It is immediately visible that time-reversal conjugates of these transitions are not identical with neither their CP- nor CPT-conjugates which is crucial for independence of the test.
\begin{table}[h]
  \centering
  \begin{tabular}{r|cl|cl}
    &  Transition &  & $\mathcal{T}$-conjugate &  \\\hline
    1 &  $\kaon \to \Kp$ & $(\ell^-,\pi\pi)$ & $\Kp \to \kaon$ & $(3\pi^0, \ell^+)$\\
    2 &  $\kaon \to \Km$ & $(\ell^-,  3\pi^0)$  &$\Km \to \kaon$ & $(\pi\pi,\ell^+)$\\
    3 &  $\akaon \to \Kp$ & $(\ell^+,\pi\pi)$ & $\Kp \to \akaon$ & $(3\pi^0, \ell^-)$\\
    4 &  $\akaon \to \Km$ & $(\ell^+,3\pi^0)$ & $\Km \to \akaon$ & $(\pi\pi, \ell^-)$
  \end{tabular}
  \caption{Possible transitions between flavour and CP-definite states and their time-reversal conjugates. For each transition a time-ordered pair of decay products which identifies the respective states is given.}
  \label{ag:tab:states}
\end{table}
\section{Observables of the test}
\begin{figure}[h]
  \centering
  \includegraphics[width=0.7\textwidth]{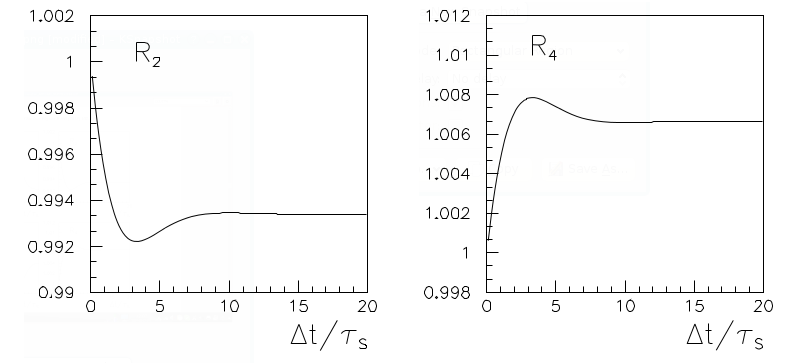}
  \vspace{-0.5cm}
  \caption{Simulated behavior of the probability ratios expected for 10$fb^{-1}$ of KLOE-2 data. The figure was adapted from\cite{Bernabeu:2012nu}.}
  \label{ag:fig:ratios}
\end{figure}

For each of the transitions from Table \ref{ag:tab:states} occurring in time $\Delta t$ and its time-reversal conjugate a time-dependent ratio of probabilities can be defined as an observable of the $\mathcal{T}$ symmetry test. In the region where high statistics is expected at KLOE-2, however, two of them are important for the test:
\begin{align}
  {R_2(\Delta t)} & = {\frac{P[\kaon(0) \to \Km(\Delta t)]}{P[\Km(0) \to \kaon(\Delta t)]} } \; \sim \; \frac{\mathrm{I}(\ell^-,3\pi^0;\Delta t)}{\mathrm{I}(\pi\pi,\ell^+;\Delta t)}, \\
  {R_4(\Delta t)} & = {\frac{P[\akaon(0) \to \Km(\Delta t)]}{P[\Km(0) \to \akaon(\Delta t)]} } \; \sim \; \frac{\mathrm{I}(\ell^+,3\pi^0;\Delta t)}{\mathrm{I}(\pi\pi,\ell^-;\Delta t)}.
\end{align}
These quantities can be measured experimentally through numbers of events with certain pairs of decays occurring in time difference $\Delta t$. A deviation of these ratios from 1 would be an indication of $\mathcal{T}$ symmetry violation. Bernabeu \textit{et al.} have simulated the behaviour of these ratios expected at KLOE-2 for 10$fb^{-1}$ of data\cite{Bernabeu:2012nu} (Figure \ref{ag:fig:ratios}). At KLOE-2 the asymptotic region of $R_2$ and $R_4$ can be observed where their theoretical behaviour may be expressed as:
\begin{align}
  R_2(\Delta t) & \stackrel{\Delta t \gg \tau_s}{\longrightarrow} 1-4\Re\epsilon, \\
  R_4(\Delta t) & \stackrel{\Delta t \gg \tau_s}{\longrightarrow} 1+4\Re\epsilon,
\end{align}
where $\epsilon = (\epsilon_S+\epsilon_L)/2$ is a T-violating parameter\cite{Bernabeu:2012nu}.

\section{Reconstruction of events for the test}
The $\mathcal{T}$ symmetry test requires reconstruction of the processes with $\Ks\to\pi\pi,\;\Kl\to\pi^{\pm}\ell^{\mp}\nu$ and $\Ks\to\pi^{\pm}\ell^{\mp}\nu,\;\Kl\to 3\pi^0$ pairs of decays. While for $\Ks\to\pi\pi$ the $\pi^+\pi^-$ final state can be chosen to take advantage of good vertex and momentum reconstruction from charged pion tracks in the KLOE drift chamber, the $\Kl\to3\pi^0\to6\gamma$ decay reconstruction is a challenging task. This process only involves neutral particles resulting in the calorimeter clusters from six $\gamma$ hits being the only recorded information. Moreover, this decay has to be reconstructed in cases where the partner $\Ks$ decays semileptonically and the missing neutrino prevents the use of kinematic constraints to aid $\Kl\to3\pi^0$ reconstruction. Therefore, this process requires independent reconstruction.

\section{The $\Kl \to 3\pi^0 \to 6\gamma$ decay vertex reconstruction}
The aim of the new reconstruction method is to obtain the spatial coordinates and time of the $\Kl$ decay point by only using information on electromagnetic calorimeter clusters created by $\gamma$ hits from $\Kl \to 3\pi^0 \to 6\gamma$. Information available for $i$-th cluster includes its spatial location and recording time $(X_i,Y_i,Z_i,T_i)$. The problem of localizing the vertex is then in its principle similar to GPS positioning and can be solved in a similar manner.
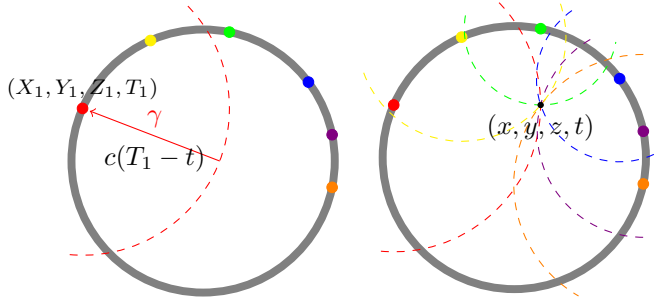
\begin{figure}[h]
  \centering
  \begin{tikzpicture}[scale=0.35]
    \coordinate (clu1) at (-4.58,2);
    \coordinate (clu2) at (1,4.90);
    \coordinate (clu3) at (4,3);
    \coordinate (clu4) at (4.90,-1);
    \coordinate (clu5) at (-2,4.58);
    \coordinate (clu6) at (4.90, 1.0);
    \coordinate (vertex) at (1.0,2.0);
    \node [] (upperphantom) at (-6,5) {};
    
    \draw[gray, line width=3pt] (0,0) circle (5);
    \draw[red,fill] (clu1) circle (0.2);
    \draw[green,fill] (clu2) circle (0.2);
    \draw[blue,fill] (clu3) circle (0.2);
    \draw[orange,fill] (clu4) circle (0.2);
    \draw[yellow,fill] (clu5) circle (0.2);
    \draw[violet,fill] (clu6) circle (0.2);
    
    \node[black,anchor=south] at (clu1) {\scriptsize$(X_1,Y_1,Z_1,T_1)$};
    \draw[red,rotate=85, dashed, thin] (clu1)+(-5.58,0) arc (180:320:5.58);
    
    \draw[thin, red,->] (0.63, 0) -- (-4.3,1.9) node[midway, above] {$\gamma$} node[midway, below, black] {\footnotesize $c(T_1-t)$};
  \end{tikzpicture}
  \begin{tikzpicture}[scale=0.35]
    \coordinate (clu1) at (-4.58,2);
    \coordinate (clu2) at (1,4.90);
    \coordinate (clu3) at (4,3);
    \coordinate (clu4) at (4.90,-1);
    \coordinate (clu5) at (-2,4.58);
    \coordinate (clu6) at (4.90, 1.0);
    \coordinate (vertex) at (1.0,2.0);
    \node [] (upperphantom) at (-6,5) {};
    
    \draw[gray, line width=3pt] (0,0) circle (5);
    \draw[red,fill] (clu1) circle (0.2);
    \draw[green,fill] (clu2) circle (0.2);
    \draw[blue,fill] (clu3) circle (0.2);
    \draw[orange,fill] (clu4) circle (0.2);
    \draw[yellow,fill] (clu5) circle (0.2);
    \draw[violet,fill] (clu6) circle (0.2);
    
    \draw[red,rotate=85, dashed] (clu1)+(-5.58,0) arc (180:320:5.58);
    
    \draw[thin, green,rotate=-10,dashed] (clu2)+(-2.90,0) arc (180:360:2.90);
    \draw[thin, blue,rotate=-55,dashed] (clu3)+(-3.16,0) arc (180:360:3.16);
    \draw[thin, orange,rotate=-90, dashed] (clu4)+(-4.92,0) arc (180:330:4.92);
    \draw[thin, yellow,rotate=20,dashed] (clu5)+(-3.96,0) arc (180:360:3.96);
    \draw[thin, violet,rotate=-60, dashed] (clu6)+(-4.03,0) arc (180:330:4.03);
    \draw[thin, black, fill] (vertex) circle (0.1) node[below]{\footnotesize $(x,y,z,t)$};
  \end{tikzpicture}
  \caption{A scheme of $\Kl\to3\pi^0\to6\gamma$ vertex reconstruction in the section view of KLOE-2 calorimeter barrel (grey circle). Colored dots denote clusters from $\gamma$ hits. Left: a set of possible origin points of a $\gamma$ which created a cluster is a sphere centered at the cluster (red dashed line) with radius parametrized by kaon flight time $t$. Right: intersection point of such spheres for all $\gamma$ hits is the $\Kl\to3\pi^0\to6\gamma$ decay point.}
  \label{ag:fig:rec}
\end{figure}

For each cluster a set of possible origin points of the incident $\gamma$ is a sphere centered at the cluster with radius parametrized by an unknown $\gamma$ origin time~$t$ (Figure \ref{ag:fig:rec}, left). Then, definition of such sets for all available clusters yields a system of up to six equations:
\begin{equation}
  (T_i-t)^2c^2 = (X_i-x)^2+(Y_i-y)^2+(Z_i-z)^2 \quad  i=1,\ldots,6,
\end{equation}
with the unknowns \textit{x,y,z} and \textit{t}.
It is then easily noticed that the $\Kl \to 3\pi^0 \to 6\gamma$ vertex is a common origin point of all photons which lies on an intersection of the spheres found as a solution of the above system (Figure \ref{ag:fig:rec}, right). At least 4 clusters are required to obtain an analytic solution although additional two may be exploited to obtain a more accurate vertex numerically.

It is worth noting that this vertex reconstruction method directly yields kaon decay time in addition to spatial location which is useful for time-dependent interferometric studies such as the $\mathcal{T}$ symmetry test.
\section*{Acknowledgments}
This work was supported in part by the Foundation for Polish Science through the MPD programme and the project HOMING PLUS BIS/2011-4/3; by the Polish National Science Centre through the Grants No. {0469/B/H03/2009/37}, {0309/B/H03/2011/40}, {2011/03/N/ST2/02641}, {2011/01/D/ST2/00748},\linebreak {2011/03/N/ST2/02652}, {2013/08/M/ST2/00323} and by the EU Integrated Infrastructure Initiative Hadron Physics Project under contract number RII3-CT- 2004-506078; by the European Commission under the 7th Framework Programme through the \textit{Research Infrastructures} action of the \textit{Capacities} Programme, Call: FP7-INFRASTRUCTURES-2008-1, Grant Agreement No. 227431.

%
\end{document}